\documentstyle[11pt,aaspp4]{article}     

\slugcomment{Accepted for publication in the {\it Ap. J.}}
\lefthead{Courteau and Rix}
\righthead{Maximal Disks and the TF Relation}

\begin{document}



\font\hugemath=cmsy10 scaled \magstep1
\def\Sum{{\hugemath\Sigma}}


\def \Rx        {R_{exp}}
\def \V22{V_{2.2}}
\def \r22{R_{disk}}
\def \partialvr {\partial\log{\V22} \thinspace / \thinspace\partial\log{\Rx}}
\def \Br        {B-r}
\def \BI        {B-I}
\def \Mr        {M_r}
\def \MI        {M_I}
\def \Reff      {R_{eff}}
\def \rc        {r_{\rm c}}
\def \gtorder{\mathrel{\raise.3ex\hbox{$>$}\mkern-14mu
             \lower0.6ex\hbox{$\sim$}}}
\def \ltorder{\mathrel{\raise.3ex\hbox{$<$}\mkern-14mu
             \lower0.6ex\hbox{$\sim$}}}

\def\reminder#1{\vskip 0.5truecm\noindent{\sl #1}}
\def \xx{{\bf ????}}

\def \etal {{\it et al.~}}
\def \eg {{\it e.g.,~}}
\def \ie {{\it i.e.,~}}
\def \vs {{\it vs~}}
\def \cf {{cf.}}

\def\o{\overline}
\def\u{\underline}
\def\t{\tilde}

\def\asec{^{\prime\prime}}
\def\kms{\ifmmode {\rm \, km \, s^{-1}} \else $\rm \,km \, s^{-1}$\fi}  
\def\ksmpc{\ifmmode{\rm km}\,{\rm s}^{-1}\,{\rm Mpc}^{-1}\else km$\,$s$^{-1}\,$Mpc$^{-1}$\fi}
\def\pc{{\rm\,pc}}
\def\kpc{{\rm\,kpc}}
\def\Mpc{{\rm\,Mpc}}
\def\dnsigma  {$D_n$-$\sigma$}
\def\lsol     {$L_\odot$}
\def\Msol{\ifmmode M_\odot\else$M_\odot$\fi}
\def\onesigma{$1$-$\sigma$}
\def\dline{\noalign{\hrule height1pt\vskip 2pt\hrule height 1pt\vskip 4pt}}
\def\bline{\noalign{\vskip 5pt\hrule height1pt}}
\def\pp{\par\yskip\noindent\hangindent 0.4in \hangafter 1}
\def\reference#1#2#3#4 {\pp#1, {\it#2}, {\bf#3}, #4.}
\def\yskip{\penalty-50\vskip3pt plus3pt minus2pt}
\def\yyskip{\penalty-100\vskip6pt plus6pt minus4pt}
\def \littlemm{\ifmmode{\scriptscriptstyle m }
     \else{\hbox{$\scriptscriptstyle m $ }}\fi}
\def \topemm{\raise .9ex \hbox{\littlemm}}
\def \magpoint{\hbox to 2pt{}\rlap{\hskip -.5ex \topemm}.\hbox to 2pt{}}
\def\deg {$^\circ$}

\def\fr#1#2{{#1\over#2}}
\def\onetwo{{\textstyle {1 \over 2} \displaystyle}}
\def\threetwo{{\textstyle {3 \over 2} \displaystyle}}
\def\alphatwo{{\textstyle {\alpha \over 2} \displaystyle}}
\def\quarter{{1\over 4}}
\def\piby2{{\pi \over 2}}
\def\threeh{{3\over 2}}
\def\fiveh{{5\over 2}}
\def\d{{\rm d}}
\def\VP{{\rm VP}}


\def\spose#1{\hbox to 0pt{#1\hss}}
\def\lta{\mathrel{\spose{\lower 3pt\hbox{$\sim$}}
    \raise 2.0pt\hbox{$<$}}}
\def\gta{\mathrel{\spose{\lower 3pt\hbox{$\sim$}}
    \raise 2.0pt\hbox{$>$}}}

\def\hub      {$H_{\hbox{\scriptsize 0}}$}
\def\hunit    {\kms~Mpc$^{\hbox{\scriptsize -1}}$}
\def\chisqr   {$\chi^2$}
\def\chidof   {$\chi^2_\nu$}
\def\ion#1#2{#1$\;${\small\rm{#2}}\relax}
\def\halpha   {{H$\alpha$\ }}
\def\ha       {H$\alpha$\ }
\def\hi       {\ion{H}{I}\ }
\def\hii      {\ion{H}{II}\ }


\title{Maximal Disks and the Tully-Fisher Relation}

\author{St\'ephane Courteau\altaffilmark{1},
        Hans-Walter Rix,\altaffilmark{2,3,4}}

\altaffiltext{1}{NRC/HIA, Dominion Astrophysical Observatory, 5071 W. Saanich Rd, Victoria, BC, V8X 4M6}
\altaffiltext{2}{Steward Observatory, University of Arizona, Tucson, AZ 85721}
\altaffiltext{3}{Max-Planck-Institut f\"ur Astrophysik,
                 Karl-Schwarzschild-Strasse 1, Garching 87540, Germany}
\altaffiltext{4}{Alfred P.~Sloan Fellow}


\begin{abstract}

We show that for luminous, non-barred, high-surface brightness (HSB) 
spirals the Tully-Fisher (TF) relation residuals can be used to estimate
the relative mass contributions of the stellar disk and the dark halo 
at the peak of the disk rotation, near $2.2$ exponential scale lengths.
For ``maximal disks", a large fraction ($0.85\pm0.1$) of the total
rotational support, $\V22$, at such radii should arise from their stellar 
mass.  Therefore the disk size, or surface-brightness, should be a significant 
additional parameter in the TF relation.  At a given absolute luminosity,
$M_r$, more compact disks (as measured by the disk scale length $\Rx$) 
should have higher rotation speeds, $\V22$.  Using a well-defined sample 
of late-type spirals, 
deviations, $\Delta\log\V22$ and $\Delta\log\Rx$, from the mean relations, 
$\V22(M_r)$ and $\Rx(M_r)$, are not significantly correlated.  The case of
$\partialvr=-0.5$ expected for a maximal disk is ruled out for the majority 
of these HSB galaxies.

We model adiabatic infall of varying amounts of luminous matter into dark 
matter halos to explore the range of possible values for $\partialvr.$  
From this, we find that the TF relation residuals require a mean value of
$V_{disk} \sim 0.6 V_{tot}$, fairly insensitive to the details of 
the initial dark matter halo and to the presence of a bulge. 
This translates to $M_{halo} \sim 0.6 M_{total}$ within $2.2 R_{exp}$,
or roughly twice more dark matter
in the inner parts of late-type spirals than previously accounted for
by maximum disk fits.  We show that any stellar population differences 
between disks of different scalelengths lead to lower values of 
$V_{disk}/V_{tot}$.  Our result is independent of the shape of the 
luminosity profile and relies only on the assumption of adiabatic contraction 
and that the dark matter halo rotation rises in the central parts of the galaxy.
Sub-maximal disks establish a natural continuity between HSB and low-surface 
brightness (LSB) galaxies which appear to be completely dark 
matter dominated even in their inner regions.
\end{abstract}





\section{Introduction}

Over the last twenty years, much effort has been invested in searching
for a ``third parameter'' to reduce the scatter in the two-parameter
Tully-Fisher relation (TFR) between the stellar luminosity and the rotation speed
of spiral galaxies (\eg Strauss \& Willick 1996, Courteau 1997; hereafter C97, and 
references therein).
For late-type spiral galaxies, no other structural parameter (such as the size,
light concentration, or mean surface brightness of the galaxy) appears to lead to a 
significant reduction in scatter.  In short, the stellar luminosity alone seems to 
determine the rotation speed (which we will define more precisely below) of a spiral 
galaxy.

It is common to model the rotation curves of individual galaxies with multi-component
mass models, usually consisting of bulge, disk and dark halo components (\eg Broeils
1992, hereafter B92; van der Kruit 1995, hereafter vdK95). 
The mass components representing the bulge and the disk are usually derived from 
the deprojected light-distribution, assuming a constant mass-to-light ratio, to be 
determined by the data.  The assumed constancy of $(M/L)_{disk}$ across a given galaxy is 
often justified on the basis of modest radial 
color gradients (de Jong 1995) and the small 
changes of vertical scale heights as a function of radius in late-type spirals 
(van der Kruit 1988, de Grijs \& Peletier 1997).  
Especially for data
taken in the redder bands this approximation should be valid.  
With only a single value of $(M/L)_{disk}$, analytical 
galaxy formation models can
explain simultaneously the properties of HSB and LSB galaxies, like the slope and
zero-point of the TF relation, very well (Dalcanton \etal 1997, Mo, Mao, \& White
1998).  This provides further support for the constancy of $(M/L)_{disk}$ among 
disk galaxies.

Dark matter halos in rotation curve decompositions have been modeled
traditionally as isothermal spheres
with a homogeneous core.  Because these multi-component fits to the rotation 
curve data are not unique, it appeared sensible to add a constraint,
such as the ``maximum disk" or ``minimum dark matter" solution (Sancisi \& 
van Albada 1985, Carignan \& Freeman 1985), to obtain a unique solution.
These fits imply the highest possible value for $(M/L)_{disk}$, typically ranging
from 3 to 5 in the $r$ band (Broeils \& Courteau 1997; hereafter BC97) and, 
in turn, lead to larger core radii for the halo.  ``Maximum disk'' fits
have been very successful, matching almost all \hi and \ha rotation curve data (B92, 
Buchhorn 1992, BC97), reproducing the small 
scale features of optical rotation curves well (Freeman 1993), and satisfying 
constraints on spirals arms for non-interacting galaxies (Athanassoula \etal 1987). 
Maximum disk models imply that for bright HSB Sb-Sc spirals, the stellar disk 
and bulge are the dominant mass components inside two exponential scale lengths.
These galaxies are typically HI-deficient in their inner parts and the mass of 
the neutral gas is negligible (Broeils 1992). 

An operational definition for the maximum disk hypothesis is that the stellar disk
provides $85\% \pm 10\%$ of the total rotational support of the galaxy at 
$R_{disk}=2.2h$ (Sackett 1997).  The lower boundary accounts for bars and massive 
bulges which reduce the contribution of the disk to the overall rotation curve.  
The upper boundary of this ratio is kept at 95\% in order to prevent hollow cores 
in the dark halo.  Still, the validity of the maximum disk hypothesis has yet to 
be demonstrated unequivocally in external galaxies because of our inability to 
constrain $(M/L)_{disk}$, independent of mass models.  

To illustrate this point, we have computed mass decompositions for 10 late-type spirals 
with extended \hi rotation curves previously modeled with an isothermal halo and 
maximum disk by Broeils (1992, p. 246).  We have repeated Broeils' modeling, but 
with a cosmologically motivated halo 
(Navarro, Frenk, \& White 1996; hereafter NFW) and two values of $(M/L)_{disk}$
corresponding to 
a maximal disk and lighter disk with $V_{disk}/V_{total} = 0.6$.  The data-model 
residuals are shown in Figs.~1 \& 2.  Details regarding the fitting procedure will 
be given elsewhere.  Clearly, one cannot discriminate in favor of any one mass model 
based on the quality of these multi-component fits; the observed \hi rotation curves 
are reproduced equally well by each fit.  A similar result was also found by BC97
who modeled 290 \ha rotation curves with $r$-band surface brightness profiles 
and found equally good fits using either a pure disk or a pure halo.  This latter 
argument is however trivial given the limitations to constrain fitting parameters 
from shallow rotation curves.  Note also that the particular shape of the NFW 
halo does not play a role in constraining $V_{disk}/V_{total}$; this is because of 
the strong covariance that exists between the concentration parameter of the NFW 
halo and $(M/L)_{disk}$.  This is true of any other halo model with a core
parameter.  An example of such fitting is found in Navarro (1997, p. 408) 
for NGC 3198. 

A related argument in support of maximum disks is that the luminous matter 
alone can reproduce the overall shape and small scale features of optical 
rotation curves (Buchhorn 1992, Freeman 1993).  
The small-scale features are almost certainly the result of spiral wave 
streaming and offer weak support for maximum disks.  This is confirmed by
matching the bumpy features of the rotation curve with similar features
in major-axis luminosity cuts.  The match to the overall shape of the 
rotation curve is met just as well by pure halo or pure disks models (BC97).

In principle, there exists a technique to constrain $(M/L)_{disk}$
uniquely, but it  is difficult to implement in practice, and it still
requires statistical arguments.
For pure exponential disks, the vertical (azimuthal) velocity dispersion 
and scale height of the stellar disk are linked to its 
mass density, independent of the halo mass (van der Kruit \& Searle 1981, Bahcall \& 
Casertano 1985).  If the disk scale length is also known, one can directly infer the mass
of the disk.  The measurement of velocity dispersions in stellar disks away from 
the central bulge and disk inhomogeneities is however difficult
for large samples.  The rapid surface brightness decline in stellar disks
requires that absorption profiles beyond 1-2 disk scale lengths be
measured with 8-10m class telescopes.  This technique also 
requires that galaxy properties be matched statistically since disk scale heights 
and scale lengths cannot be measured in face-on and edge-on galaxies simultaneously.  
One attempt at this technique has been published so far (Bottema 1993) which 
suggests, based on a sample of 12 bulge-less systems and some assumptions 
about their stellar populations, that late-type spirals are 
sub-maximal with $V_{disk}/V_{total}$ equal to $63\% \pm 10\%$ at $\r22.$

Locally, most investigations of the Galactic mass density near the Sun support 
the notion of a significantly sub-maximal Milky Way disk with 
$V_{disk}/V_{total} \approx 0.5$ at $R = R_0$ (Kuijken 1995 and references therein).  
Sackett (1997) however argues that new or revised structural data for the Milky Way 
would favor a maximal Galactic disk.  A crucial element in these models is the
precise contribution of the massive central bar (\eg Zhao, Spergel, \& Rich 1995) 
or elongated bulge (Kuijken 1995) to the dynamical support in the inner parts of 
the Galaxy, which is quite uncertain at present.
The uncertainty in the disk scale length is also an issue; a 25\% difference
above or below the nominal value of 3.5 kpc is sufficient to lend a sub-maximal
result or not.  Published values for the disk scale length of the Milky Way 
currently differ by more than 30\% (Sackett 1997). 


Numerical and analytical models of disk formation in a dissipationless 
dark matter halo also predict, for realistic total fractions of baryonic
to dark matter, that spiral disks should be far from maximal 
(Blumenthal \etal 1986, Barnes 1987, Flores \etal 1993, Moore 1994, 
NFW, Dalcanton \etal 1997, Mo, Mao, \& White 1998, Navarro 1998).  


In this paper we explore whether the Tully-Fisher relation, and its seemingly 
sole dependence on stellar luminosity, is consistent with ``maximal'' disks.
The basic idea of our test is best described by a
{\it Gedankenexperiment}:
Consider a pure exponential disk galaxy of a given luminosity,
mass-to-light ratio, and disk scale length.
In the absence of dark matter or bulge,
its rotation curve is given by (Freeman 1970), 
\begin{equation}
  v_c^2(R)= 4\pi G \Sigma_0 \Rx y^2 \bigl [ I_0(y)~K_0(y) - I_1(y)~K_1(y) \bigr ],
\end{equation}
where, $G$ is the gravitational constant, $\Sigma_0$ the central surface 
brightness, $\Rx$ the disk exponential scale
length, $y\equiv R / (2\Rx)$, and $I_i(y)$ and $K_i(y)$ are the modified Bessel
functions of the first and second kind. The disk rotation curve
will peak at $\r22 = 2.15\Rx$ at a value of $\V22$. For disks of finite
thickness (say, $h/R_{exp}=0.2$)
the rotation curve will have a very similar shape but a $\sim 5$\% lower peak
(\eg Casertano 1983); this will slightly affect the ``shape"
term in square brackets in Eq. (1), but leave the subsequent scalings untouched. 

Imagine ``compressing" the same disk to a somewhat smaller scale length.
The more compacted disk will peak at a higher rotation velocity,
and dimensional analysis of Eq. 1 yields 
\begin{equation}
  \partialvr=-0.5,
\end{equation}
at fixed $M_{disk}$. Thus, if dark matter is negligible inside 2.2$R_{exp}$, a 
20\% change in disk scale length -- at a given disk mass -- should translate into 
a 10\% change of $\V22$, hence into a $\sim 30$\% offset in luminosity from 
the mean TFR.
Such an offset should be easily detectable in a statistical sense, given
the size and quality of TFR samples currently available.
Of course, this effect will be altered by the presence of a dark matter halo
and of a bulge, and we will devise a sequence of models
that predicts $\partialvr$ for a wide range of disk-to-halo
mass ratios.

Even though in practice such a test requires to substitute
$L_{disk}$ for $M_{disk}$, our test does not rely on any
assumption that $(M/L)_{disk}(R)=$const; it merely requires 
that $(M/L)_{disk}$ be independent of surface brightness and 
that any gradient in $(M/L)_{disk}(R)$ for comparable disks 
be self-similar.

In \S2 of this paper, we determine the mean value of $\partialvr$
for a set of well defined samples of luminous ($\sim L_*$), non-barred HSB galaxies.
In \S3, we compare this observed value to the predictions from a number of 
disk$-$halo models and examine the effects due to color gradients. This 
enables us to estimate what fraction of the rotation speed, $V_{disk}/V_{tot}$
at $\r22$ arises from the stellar disk.   We discuss our results in \S4 and 
offer a brief conclusion in \S 5. 

\section{Determination of $\partialvr$}

\subsection{Using Two Data Samples}

To date, two large optical TF data sets have been published, from which 
we can determine 
$\V22$ for many hundreds of non-barred HSB galaxies.  All galaxies are
selected from the field, where dark matter halos have likely not 
overlapped or recently merged.  We first use the 
collection of 306 Sb-Sc northern galaxies from Courteau and Faber (Courteau 1992; 
hereafter referred to as CF). The CF galaxies were selected from the UGC catalog 
using the following criteria: $m_B < 15.5$, blue major axis diameter $\leq 4^\prime$, 
inclination between 55\deg\ and 75\deg, no bar, and no apparent
interaction.  
The velocities $\V22$ were modeled via parametric fitting of the H$\alpha$ 
rotation curves (C97) and the CCD magnitudes and disk scale lenghts were measured 
in the Gunn $r$ band (Courteau 1996). 
Specific details about galaxy selection can be found in Courteau (1992, 1996).  
We use a sub-sample of the CF catalog in a region of Quiet Hubble Flow (QHF) to 
minimise the effects of streaming and local infall motions on distance estimates
(Courteau \etal 1993).  
For this analysis, the CF QHF sub-sample is further restricted to rotation curves 
that extend beyond 
$2.5 \Rx$ (so that $\V22$ is defined) and to $M_r<-20$.  Only the brighter
galaxies ($\sim L_*$) are retained where the dynamical contributions from the 
stellar disk are likely to be largest. All other cases, expect perhaps for 
bar-dominated galaxies, presumably yield a lower fraction of visible-to-total 
matter at $\r22$.
The $B-r$ color is also computed from the RC3 B-band and 
CF $r$-band magnitudes are corrected for Galactic and internal extinction.  
After all culling, 124 CF/QHF galaxies are kept for final analysis.  

We also use the collection of 1355 southern late-type spirals from Mathewson, Ford, 
\& Buchhorn (1992; hereafter MAT).  Peculiar motions are not as conspicuous in 
this sample and no spatial culling is applied.  Excluding
or retaining galaxies in the vicinity of the Great Attractor
does not change the final result.
As above for the CF sample, we restrict the total MAT sample to $R>2.5 \Rx$ and 
$\MI<-20.4$, and require galaxies to have a measured $B-I$ color and I-band scale 
length.  This yields a final sample of 406 
MAT galaxies.  The $I$-band magnitudes and $B-I$ colors were corrected 
for internal and Galactic extinction as described in Willick \etal (1997). 
The fitting of $\V22$ for the MAT galaxies is described in C97.  
The disk scale lengths, from Byun (1995), were extracted from 2D bulge/disk 
decompositions of the galaxy images. 

Five galaxies overlap between the CF and MAT samples.  With those, we 
can verify that our CF scalelengths and those measured by Byun agree to
within 10\%.  The typical uncertainty in published scalelengths for a 
given galaxy is about 23\% (Knapen \& van der Kruit 1991).  The matching 
rotational amplitudes $\V22$ also agree very well, within 5\% (C97). 

It is of interest to verify how adequately the surface density of disk galaxies 
is described by a pure exponential profile.  Using this information, we can
show that our final result does not depend on any assumption about the shape 
of the luminosity profile of the disk. 

We compute the fraction of ``exponential light'' at a given radius as the ratio
of the integrated light, based on the disk exponential fit, to the 
total observed light inside that radius.  We measure two separate ratios at 
$\r22$ and at ``infinity.''  The total magnitude measured at infinity includes 
all the light from the galaxy plus a small extrapolation of the disk profile 
(typically $0\magpoint02$) from its observed edge to infinity. 

A disk is deemed ``purely exponential'' if the ratio of ``exponential'' to 
observed light at a given radius does not deviate by more than 20\% from unity.
In Fig.~3, we show exponential light ratios computed at 
$\r22$ and infinity for all 306 CF galaxies.
More than three-quarters of the light profiles deviate from the idealized 
exponential disk model, some by more than 50\%.
Galaxies left of the peak of the histogram have typically a larger
bulge or can be described by a Freeman Type I luminosity profile (Freeman 1970),
whereas those on the right side correspond to a Freeman Type II profile
or will show a truncated disk.  Based on the criterion above, 85 CF galaxies,
or 22\% of the total CF sample, qualify as pure exponential disks.
This ratio is closer to 18\% for all MAT galaxies. 

The TF distribution of all CF galaxies is shown in Fig.~4.  The two 
distributions with circles and squares in each panel use absolute magnitudes 
{\it measured} at $\r22$ and infinity respectively.  The solid line is 
the QHF TF fit from C97 and the dashed line is offset from that fit by
0\magpoint5.  This magnitude difference is that expected between the 
light measured at $\r22$ and at infinity for a pure exponential disk.
The TF slopes are statistically equivalent 
and the zero-points differ by 0\magpoint62 $\pm$ 0\magpoint05, a 
little over the exponential disk value.
The bottom panel shows the same distributions for pure exponential 
disks only; here, the TF zero-point offset is exactly 0\magpoint5, 
as expected by construction.  The TF scatter for the QHF galaxies
increases from 0\magpoint34 when total extrapolated magnitudes are used 
to 0\magpoint40 if the light interior to $\r22$ is used.
As expected, there is no dependence of the TF scatter on the choice of 
magnitudes for the pure exponential galaxies ($\sigma_{\rm{TF}}=0\magpoint39$ 
in either case).
%
%
%
%
In all cases, whether one considers the full CF sample or only 
the distribution of pure exponential disks, the TF slopes are equivalent
within their statistical error (independent of the choice of magnitude),
meaning that dynamically, all late-type spirals behave as exponential 
disks\footnote{It also implies that the typical rotation 
curve of a late-type spiral has reached turnover for $R>R_{disk}$.
(C97).}.  

The data used in this study are: $\V22$, $\Rx$, $M_r$ or $M_I$, a $B-r$
or $B-I$ color term, distance of the galaxy in Mpc (see C97), and the 
fraction of exponential light in a galaxy.  Distance-dependent quantities 
assume \hub$=70$\hunit\ as in C97. 

\subsection{Fitting the Mean Parameter Relations}

A galaxy's rotation speed, size, and color correlate with its absolute luminosity; 
these correlations are known as the Tully-Fisher relation (Tully \& Fisher 1977), 
Freeman's law (Freeman 1970), and the color-magnitude relation (Tully \etal 1982),
and are shown in Fig.~5 for our two samples.  At present, we want to ask how various 
galaxy properties correlate {\it at a given absolute luminosity}.  Hence
we need to fit and remove the trends of $\V22$, $\Rx$, and color ($\Br$ or $\BI$)
with absolute magnitude ($\Mr$ or $\MI$ for each sample).  These three relations 
were fit by the functional form 
$f_{\bar{y},\alpha_y}(y) = \bar{y} + \alpha_y (\Mr - \bar{\Mr})$, 
where $y$ stands for $\log\V22$, $\log\Rx$, and $\Br$, respectively, $\Mr$ is the 
absolute magnitude of the CF galaxies, and the upper bar denotes a median value.
The same notation applies for the MAT sample with the appropriate terms replaced.

To achieve a robust fit to the data $\bigl [ y(i),\Mr(i)\bigr ]$,
we varied $\bar{y}$ and $\alpha_y$ to minimize 
the sign of the data$-$model deviation
(Press \etal 1992, \S 15.7).  These fits are shown in Table 1.  Standard least-squares 
fits give essentially the same solutions. 

\begin{table}[th]
\centering
\tablenum{1}
\begin{tabular}{l c c c}
\multicolumn{4}{c}{Table 1. Parameter Correlations} \\ [2pt] \hline\hline 
 & & & \\ [-12pt]
& CF/QHF galaxies \ (N=124)&& CF pure exp. disks \ (N=28) \\ [1pt] \cline{2-2} \cline{4-4} 
 & & & \\ [-13pt]
$\log\V22(\Mr)$ & $2.278 - 0.136\cdot (\Mr + 21.30)$ && $2.272 - 0.125\cdot (\Mr + 21.30)$ \\ [1.2pt]
$\log\Rx(\Mr)$  & $0.674 - 0.158\cdot (\Mr + 21.30)$ && $0.686 - 0.142\cdot (\Mr + 21.30)$ \\ [1.2pt]
$\bigl[\Br\bigr](\Mr)$ & $0.890  - 0.238 \cdot (\Mr + 21.30)$ && $0.890  - 0.200 \cdot (\Mr + 21.30)$ \\
 [3pt] \hline 
 & & & \\ [-12pt]
& MAT galaxies \ (N=406)&& MAT pure exp. disks \ (N=73) \\ [1pt] \cline{2-2} \cline{4-4} 
 & & & \\ [-13pt]
$\log\V22(\MI)$ & $2.244 - 0.134\cdot (\MI + 21.59)$ && $2.261 - 0.133\cdot (\MI + 21.72)$ \\ [1.2pt]
$\log\Rx(\MI)$  & $0.632 - 0.130\cdot (\MI + 21.59)$ && $0.619 - 0.120\cdot (\MI + 21.72) $ \\ [1.2pt]
$\bigl[\BI\bigr](\MI)$ & $1.649  - 0.192 \cdot (\MI + 21.59)$ && $1.649  - 0.222 \cdot (\MI + 21.72)$ \\ 
 [3pt] \hline\hline
\end{tabular}
\end{table}
 

The fits were applied to the sample of exponential disks only and are 
shown as solid lines in Fig.~5.  Open and closed symbols show all 
the galaxies and the restricted sample of pure exponential disks
respectively. 


\subsection{Correlations among the Residuals}

We can now define the residuals for each object $i$
as $\Delta y(i)\equiv y(i)-f_{\bar{y},\alpha_y}[y(i)]$. 
Fig.~6 shows the residuals for $\log\V22$, $\log\Rx$, 
and color plotted against each other. 
>From inspection of the plots,
it is not obvious which range of correlation slopes 
$\partial (\Delta y_1)/\partial(\Delta y_2)$ is statistically acceptable.  
We devise a robust non-parametric test by rotating the set of
residuals $\bigl (\Delta y_1(i), \Delta y_2(i) \bigr )$ by various angles 
$\theta$ to get $\bigl ( \Delta \hat{y}_1(i),\Delta \hat{y}_2(i) \bigr )$ 
and then applying a Spearman rank test (Press \etal 1992, \S 14.6)
for correlation between the quantities $\bigl (\Delta \hat{y}_1(i),
\Delta \hat{y}_2(i) \bigr)$.
The acceptable range of  $\partial (\Delta y_1)/\partial (\Delta y_2)$ 
can be calculated from the range of angles $\theta$ for which the 
$\bigl (\Delta\hat{y}_1(i),\Delta\hat{y}_2(i) \bigr )$ are {\it not} 
significantly correlated.  

Table 2 shows correlation slopes for the full and restricted 
CF/QHF and MAT samples. The error quoted for each slope gives the 
95\% confidence 
level.  

\begin{table}[th]
\centering
\baselineskip=17pt plus 1pt
\tablenum{2}
\begin{tabular}{l c c c cc }
\multicolumn{6}{c}{Table 2. Residual correlations} \\ \hline\hline
 & & & \\ [-14pt]
& \multicolumn{2}{c}{CF/QHF galaxies} & & \multicolumn{2}{c}{MAT galaxies} \\ \cline{2-3} \cline{5-6}
 & & & \\ [-13pt]
& \ \ all & \ \ exp. only && \ \ all & \ \ exp. only \\ \hline
 & & & \\ [-13pt]
$\partialvr$ &$-0.070\pm0.122$&$-0.104\pm0.137$&&$-0.180\pm0.051$&$-0.233\pm0.098$ \\ [1pt]
$\partial\log\Rx \thinspace / \thinspace \partial(color)$&$-0.073\pm0.071$&$-0.197\pm0.198$&&$-0.030\pm0.054$&$-0.147\pm0.102$ \\ [1pt]
$\partial\log\V22 \thinspace / \thinspace \partial(color)$&$\phantom{-}0.052\pm0.029$&$\phantom{-}0.011\pm0.060$&&$\phantom{-}0.050\pm0.019$&
$\phantom{-}0.027\pm 0.039$ \\ \hline\hline
\end{tabular}
\end{table}

The first line of Table 2 shows that there is tentative evidence that
$\partialvr<0$, indicating that the disk mass does play a role in
setting $V_{2.2}$. However, the case of  $\partialvr=-0.5$, 
expected for the disk-only case, is clearly ruled out. 

Further, the residual analysis shows that 
there is a slight (but not statistically significant)
tendency that redder disks (at a given
luminosity) are slightly more compact,
and that redder disks have higher $V_{2.2}$ (see Table~2).
These trends are expected in a variety of
evolutionary models for disk galaxies (\eg Firmani \&
Avila-Reese 1998).

Before proceeding with a dynamical analysis (\S 3) one should
explore the potential impact of stellar population variations
among disks of the same luminosity.
Qualitatively, one would expect redder disks to have higher
mass-to-light ratios, whether their colors are affected by dust,
greater age, or higher metallicity. Since, more compact disks
are observed to be marginally redder (Table 2),
this would imply that the derived
$\partialvr$ is lower than its ``true" value, and consequently
the observed mean value of $\partialvr = -0.19 \pm .05$ is an
{\it upper} limit to the actual dynamical contribution
from the disk stars.

To be more quantitative about 
M/L variations {\sl among} different
disks, we have
used population models (Vazdekis \etal 1996, their Table 2)
to determine how the stellar $r$-band $M/L$ varies with
color. 
As a full exploration of this question is beyond the
scope of this paper, we follow the common practice
of using single age population models to match colors that arise
from a more complex star-formation history, but
which has resulted in the same luminosity weighted stellar age.
We use population models with a mean age of $1-4$~Gyrs and 
near-solar metallicity to find that 
$\partial \log{(M/L)_r}~/~\partial [B-r] \sim 1$ if the 
color differences are attributable to age differences,
$\partial \log{(M/L)_r}~/~\partial [B-r] \sim 0.4$ if they
are attributable to metallicity differences, and 
$\partial \log{(M/L)_r}~/~\partial [B-r] \approx R_{B-r}/2.5
\sim 1$ for dust reddening, where $R_{B-r}$ is the ratio of
total $r$-band extinction to the differential extinction 
$E(B-r)$. This last quantity will depend on the details of 
the dust distribution. For the benchmark case of a simple absorption
screen with a mean Galactic extinction curve, $R_{B-r}\sim 2.5$
(Savage and Mathis, 1979).

In the context of these models, it
appears the reddening due to stellar population age differences,
and dust reddening imply the largest $(M/L)_r$ differences,
with $\Delta (B-r)\sim 0.1$ corresponding to a 
25\% increase in $(M/L)_r$.


Another potential bias on the final value of $\partialvr$ is the 
wavelength dependence on scalelengths. 
The stellar disk mass is dominated by low-mass stars which emit most of their light 
in the redder bandpasses.  Therefore, K-band scalelengths may be most representative 
of the surface density of the disk population. 
Scale lengths at red band passes are typically $\sim 15\%$ shorter
than in the blue
(de Jong 1995, Peletier \& Balcells 1997). 
The ratio between the $r$ and $I$ band scalelengths and 
K-band scalelengths is about half that value but we will test for a more extreme 
difference of 10\%.  That is, we now measure $V_{disk}$ at $r=1.90h$ and use the shorter 
scalelength as our new fiducial.  For both CF and MAT galaxies, this yields 
an 8\% increase of $\partial\log V_{disk} \thinspace / \thinspace \partial \log\Rx$.
This effect is negligible and does not impact our conclusions. 
In the following we develop a dynamical analysis under the 
assumption that {\it within our sample} the stellar populations of 
disks at the same luminosity are similar.


\section{What values of $\partial \log\V22/\partial\log \Rx$ should we expect?}

If the stellar disk were the only relevant mass component, 
we would expect $\partialvr=-0.5$, in the absence of color gradients.
If, on the other hand, the stars are only test particles, then
a difference in $R_{exp}$ simply shifts the physical radius
at which we determine $\V22$. Therefore, in this limiting case
$\partialvr$ just reflects 
$\partial\log V_{halo}/\partial\log R_{exp}$, the logarithmic gradient 
of the halo rotation curve at $2.2$ disk scale lengths. 
Note that for halos with rising rotation curves (isothermal with a
core, or NFW) one finds $\partialvr > 0$. As a consequence, one
would expect a surface brightness, or disk size, dependence in the
$M_r - \V22$ relation, both for the case of dominant and of
negligible disk mass, yet with {\it opposite sign}.
Consequently, there must be an intermediate case, where the stellar
disk is somewhat important and the surface brightness disappears.

The actual situation 
will be bracketed by these two extremes.  Below, we explore how
$\partialvr$ depends on the fraction of the circular velocity at $\r22$ 
that is attributable to the stellar disk, $V_{disk}/V_{total}$.

\subsection{Exponential Disks in NFW Halos}

We first examine a simple exponential disk embedded in dark
matter halo. For the {\it initial} dark matter halo structure,
we use a density profile $\rho_{DM}(r)\propto (r/r_s)^{-1}[1+(r/r_s)]^{-2}$
found in the collisionless CDM simulations of halo formation by NFW.
These simulations indicate that this initial structure only depends
on the total halo mass, which translates into the total stellar
mass, if the baryonic fraction is constant among halos of identical
mass, and if most cooled baryons have been converted into stars.
 
If the central accumulation of the baryons occurs 
on a longer time scale than the local dynamical time,
the dissipationless dark matter halo will contract
obeying adiabatic invariance (Blumenthal \etal 1986).  
Calculating this adjustment is easy in spherical geometry 
(\eg Flores \etal 1993, NFW), but 
requires replacement of the disk configuration
by the spherical density profile that has the same enclosed mass $M(<r)$
(see Binney \& Tremaine 1987, Eq. 2-170).  It is well known that this 
procedure leads to a somewhat incorrect rotation curve, but we are 
only interested in derivatives.
The final disk rotation curve is still computed using Eq. 1, and
the final dark matter distribution is given by
$M_{\rm halo}(r) = M_{\rm halo}(r_i) = (1 - \Omega_b) M_i(r_i)$
where $M_i(r_i)$ is the total mass within radius $r_i$ before disk 
formation (\eg NFW). 

Using a SCDM halo profile from NFW with $V_{200}=180$\kms\ and 
assuming a baryon fraction of $\Omega_b = 0.04$, we calculated the 
combined disk$-$halo rotation curves for various $(M/L)_{disk}$ and 
for $\Rx=3$~kpc, corresponding to a range of $V_{disk}/V_{total}$ 
at $\r22$.  Then we changed $\Rx$ by 5\% retaining the total disk mass
and the {\it initial} (uncontracted) halo profile, and repeated the 
calculation to estimate $\partialvr$ for each value of $V_{disk}/V_{total}$.  
Our choice of $V_{200}$ and $\Rx$ is representative of the Milky Way and 
the $L_*$ galaxies considered in this study.  We will show below that our 
result is rather insensitive to reasonable variations in our disk and halo 
parameters. Experiments with $\Delta\Rx/\Rx=10\%$ showed that the finite 
change has no significant impact on the estimate of the gradient.  In cases
where a central bulge is included (\S 3.2), the spheroid remains unchanged 
when considering the small changes in 
$\Rx$ for the gradient calculation.  An example of such a contraction is 
shown in Fig.~7.  The estimated gradients are shown as the short-dashed line 
connecting the open circles in Fig.~8. Note also that for low mass disks, 
$V_{disk}/V_{total}< 0.45$, the gradient $\partialvr$ becomes positive,
because the halo rotation curve is rising, and an increase in $\Rx$ means 
that $\V22$ is measured at greater metric radii. 

These results are insensitive to the adopted disk scale length or
rotation curve amplitude provided representative values are used.
If we choose $\Rx=2$~kpc (4 kpc) in the calculations, the value of
$\partialvr$ at a given $V_{disk}/V_{total}$ increases (decreases) in
the mean by only 0.03 (see Fig.~9) from our nominal case.  Disk scale
lengths for HSB galaxies
vary typically between 2kpc and 6kpc (Courteau 1996).  Similar modest
excursions for $\partialvr$ are found if we use different rotation curve
amplitudes.  The value of $V_{200}$ is a slow function of the optical
velocity, or $\V22$,
and is typically smaller than $\V22$ for HSB galaxies (the opposite is
true for LSB galaxies).  We have experimented with a range of halo
velocities from 160 to 200 \kms, or a variation in
$\V22$ of 150 to 300 \kms\ (depending on the cosmology).
The value of $\partialvr$ increases in
the mean by less than 0.03 per 10 \kms\ increments in $V_{200}$ for a
given disk mass and mass-to-light ratio.  This is because the shape of
the logarithmic NFW rotation curve changes little over such amplitude
increments at $\V22$.  This is also depicted in Fig.~9.  These tests
were performed with and without a bulge component (see below).  Thus
our results are only weakly dependent on the exact value of $V_{200}$
and $R_{exp}$ provided we use reasonable estimates for $L_*$ galaxies.
Our nominal halo velocity, $V_{200} = 180$ \kms\ corresponds to
$\V22 \sim 220$ \kms\ for an $\Omega=0.2$ CDM universe
(Navarro 1998).
 
Our choice of baryon fraction is consistent with the latest 
D/H abundances (Burles \& Tytler 1998, Levshakov \etal 1998) which give 
$\Omega_b h_{100}^2 = 0.017 \pm 0.002$.  The specific choice of 
$\Omega_b$ is inconsequential for this analysis; results are essentially 
unchanged if we adopt $\Omega_b = 0.04 \pm 0.02$ (\hub =70).

\subsection{Central Bulge }

Even considering the late-type spirals of our sample, the bulge
component may not be dynamically negligible. Therefore,
we repeated the above calculation including in the rotation curve
calculation a spheroid with 20\% of the disk mass and a scale
radius of 500~pc.
Such a bulge mass fraction is an upper-limit for these Sb$-$Sc
galaxies (BC97).  We repeated the same analysis as above with 
this new component in the halo contraction and rotation curve
calculation. The resulting $\partialvr$ are shown
as the solid line connecting the triangles in Fig~9.  
It is evident that even a large bulge for these galaxies makes little 
difference for this calculation at $\r22$ (see also Mo \etal 1997). 

\subsection{Isothermal Halos with Cores}

Finally, we explore how sensitive are these results to the particular 
functional form of the assumed dark matter profile. To this end, we 
repeated our calculations starting from an isothermal dark matter halo 
with an initial core radius $r_c=3\Rx$.  The resulting dependence 
of $\partialvr$ on $V_{disk}/V_{total}$ is shown by the long-dashed 
line connecting the open squares in Fig.~8.  The result is again quite 
similar to the previous two cases.  Choosing a larger core radius makes 
little difference for heavy disks ($V_{disk}/V_{total} \gtorder 0.6$), 
but leads to rotation curve shapes for very light disks that are linearly 
rising within $\sim 2\Rx$, grossly inconsistent with the observations for
these luminous, HSB disks (C97). 

\subsection{Dependence on Initial Halo Parameters}
The derivative $dlogV_{2.2}/dlogR_{exp}$ is a function of (at least)
four variables: the initial values of $V_{200}$ (assumed to be 
directly coupled to the NFW concentration parameter) and $R_{exp}$, the 
disk size, mass and disk mass-to-light ratio.  If the initial halo structure
and size of the disk were coupled, {\it i.e.} if $V_{200}$ varied 
systematically with $R_{exp}$, it would change the interpretation of 
$dlogV_{2.2}/dlogR_{exp}$.  To impact our conclusions, the denser, or 
more concentrated, halos of a given mass scale (NFW predict a 25 \% 
scatter in their concentration parameter) would need to harbor
larger than average disks.  The dark matter could aid the lower 
surface brightness stellar disks to yield the same rotation speed 
at $\V22$ as that of their higher surface brightness cousins and 
remove any surface brightness trend in the TF relation, even in the presence
of a near-maximal disk.  

There seems to be broad agreement that (at least for systems with 
small bulges, \ie little merging) the disk size is regulated by the 
initial angular momentum ({\it e.g.} Mo, Mao and White, 1998). 
For the purpose of our analysis, it suffices that there exists
a monotonic mapping between total initial angular momentum
and final disk size; this would mean that larger disks (at a 
given total baryonic mass) come from baryon/dark matter systems with 
higher initial spin parameters $\lambda$ (see also \S3.5.2 in
Mo, Mao \& White 1998).  When considering halos of the same mass scale
but with different spin-parameters $\lambda$ one might suspect that 
the halos, too, become more extended (or less concentrated) at higher 
$\lambda$, analogous to the disks.  In this case the surface mass density 
of both disk and halo would drop at higher $\lambda$, and one would 
necessarily expect an SB dependence of the TFR.

The halo concentration is found to be independent of the spin parameter 
$\lambda$ (see also \S3.5.2 in Mo, Mao \& White 1998).  Hence, at
least in the context of hierarchical formation simulations, one 
would not expect any correlation between the (initial) halo structure 
and size of the disk.  This is implicit in our derivation.  We calculate 
$\partialvr$ using halo parameters that depend only on the total mass.
Hence our conclusion is somewhat model dependent, but one which follows 
from current hierarchical scenarios.  Note that the initial halo choice 
matters least for a dominant disk, where the final halo structure in the 
inner parts is predominantly determined by the (adiabatic) contraction 
process.

We have not investigated the (random) scatter in the initial halo 
concentrations but this would merely contribute to weakening our 
model-data comparison and the power of our statistical tests and we 
believe our conclusions to be conservative in this respect.

\section{Discussion}

The mean value of $\partialvr$ in both CF and MAT samples of exponential 
disks is $-0.19 \pm 0.05$.  By taking the range of $V_{disk}/V_{total}$ 
for which our models predict $\partialvr$ in this range (Fig.~8), we find 
a mean value of $V_{disk}/V_{total} =0.61 \pm 0.07$.  This range can be 
widened if we adopt limits determined from each sample separately; in 
this case, we would have $V_{disk}/V_{total} = 0.59 \pm 0.16$.  
Therefore, our mean relation is inconsistent 
with $V_{disk}/V_{total} > 0.75$, the mimimum boundary for a 
``maximal disk'' (Sackett 1997), at the 95\% confidence level.
This statement holds for any reasonable combination of $V_{200}$, 
$R_{exp}$, the mass of the disk, and $(M/L)_{disk}$.  That is, we 
have tested for any combination of our model input parameters and 
cannot produce $V_{disk}/V_{total} > 0.75$ for $\partialvr \geq -0.3$,
the absolute maximum allowed by our global data\footnote{This 
statement applies to the data ensemble; clearly, it may be possible to 
find small sets of galaxies where $V_{disk}/V_{total}$ exceeds
0.75.}.

The need to consider the actual distribution of the light
in any dynamical measurements (\eg mass decomposition), instead
of assuming {\it a priori} an average luminosity profile, cannot be 
overemphasized (see \eg Kalnajs 1983).  We did indeed restrict 
our analysis to galaxies with a nearly pure exponential light profile.
However, this distinction turns out to matter little 
for our statistical analysis.  Inclusion of all types of galaxy
profiles would bias the value $\partialvr$ towards larger fractions of 
dark-to-total matter at $\r22$ by no more than 15\% of the actual value 
for pure exponential disks, with $V_{disk}/V_{tot} \simeq 0.55$
and a 95\% upper limit of 0.66 (Courteau \& Rix 1998). 
Inspection of Eqs. 1 and 2 shows that our analysis is insensitive
to the assumption of exponential disks: as long as the rotation
curve shape factor in Eq.~1,
$[ I_0(y)~K_0(y) - I_1(y)~K_1(y)]$, does not depend strongly on 
$\Rx$ at a fixed luminosity,  Eq. 2 holds exactly.
For the remainder of this discussion, we adopt the ratio 
$V_{disk}/V_{total} =0.6 \pm 0.1$ as our nominal result.  

Our main argument is based on possible correlations between $\V22$
and $\Rx$ which can be ruled out statistically in the presence of
the observed scatter.  As long as the sources of the scatter are
not systematically linked to the correlation of interest, their 
nature, cause or origin is inconsequential.  For example, inclination
errors and Hubble-flow distance errors are not systematically linked 
to either $\V22$ and $\Rx$ and we did not consider them explicitly\footnote{
These merely soften the confidence limits on possible correlation slopes}.
Color gradients (dust, stellar populations) are relevant sources of scatter
that might vary systematically with disk scale length.  However, as we 
saw in \S2.3, correction for color differences can only increase the 
ratio of dark-to-visible matter at $\r22$.  This is because the mean 
inferred value of $V_{disk}/V_{total}$ is an upper limit to the 
contribution of the disk mass to the rotation speed at $\r22$.  
The dynamical interpretation of this gradient is only slightly 
altered if we consider the dependence of stellar population on 
disk mass-to-light ratios (Bottema 1997, hereafter B97).  
B97 examined population differences between LSB and HSB galaxies
using crude population synthesis arguments, and showed that 
dust or metallicity gradients in spiral galaxies can account for
$\sim 15\%$ of $V_{disk}/V_{total}.$
Our own estimates in \S2.3 using the population models of Vazdekis 
\etal 1996, suggest that the impact of age and dust on $V_{disk}/V_{total}$
can be even larger with an increase of 25\% in $(M/L)_r$, or a 50\%
decrease of $V_{disk}/V_{total}$.  This would lower $V_{disk}/V_{total}$ 
to $0.3 \pm 0.2,$ for color offsets $\Delta (B-r) \sim 0.1$ from the mean.  
This value is in agreement with the data if the maximum observed color 
trends seen in Fig.~8 for $\Rx$ and $\V22$ are removed from the distribution
of $\partialvr$.  However, because colors depend strongly on the star 
formation history of the galaxy, more detailed chemical and spectral 
evolutionary models based on different galaxy formation scenarios 
(merging, extended collapse) are needed to address the interplay 
between colors and the dynamical properties of galaxies on firmer 
grounds (\cf \ \eg Firmani \& Avila-Reese 1998, Sommerville \& Primack 1998).

Arguments in favor of sub-maximal disks have been given by van der Kruit (1995) 
and a few were highlighted in \S 1.  Abiding by the definition that a disk is
``maximal'' if $V_{disk}/V_{total} = 85\% \pm 10\%$ (Sackett 1997), we have 
effectively shown that non-barred late-type spirals are, {\it on average}, 
sub-maximal at $\r22$,  and that the halo contributes
$M_{halo} \gta 0.6 M_{total}$, within  that radius.
Sub-maximal disks with $V_{disk}/V_{total} \sim 0.6$ imply $(M/L)_{disk} 
\approx 1-2$, a factor of two smaller than maximal disk values (Bottema 1993, 
BC97).  This corresponds to twice more dark matter in the inner parts of late-type 
spirals than previously accounted for by maximal disk fits. 
Evolutionary synthesis models of spiral galaxies suggest mass-to-light ratios
in the range 1-2\footnote{Although the range of $M/L$ ratios depends 
significantly on the choice of initial mass function.  Here, we assume a
Scalo IMF truncated at $0.1\Msol$ and $100\Msol$} in the $r$-band, with 
possible extremes up to 5 for very red systems (Bruzual \& Charlot 1993).
Thus, there is good agreement between these
models and our result.  The same is also true with the work of Bottema (1993) 
who combined measurements of velocity dispersion, scale length, and scale height 
of exponential disks to infer that old disks contribute at most $63\%\pm10\%$ of 
the observed maximum rotation at a given luminosity.  
Another supportive result is that of Quillen \& Sarajedini (1998) who find that 
low values of the disk $(M/L)_I$ in the range 1-2 are required to explain stable 
disks at $z\approx1$.  In order to avoid unrealistically thick disks, one must
invoke a substantial amount of dark matter in the inner parts of these
distant disks, in agreement with our study of low redshift galaxies.

Swing amplification theory (Athanassoula \etal 1987) puts a constraint
on the density of galactic disks in order to preserve m=2,3 modes
(and inhibit m=1 modes).  The low values of $(M/L)_{disk}$ observed here
are in complete agreement with the observed spiral structure of late-type 
disks.  

While our statistical argument favors sub-maximal disks in the majority of 
HSB non-barred disk galaxies, it does not completely exclude the possibility 
that some galactic disks could be maximal.  Such disks are more likely to be 
found in lighter or less centrally concentrated halos or equivalently, 
systems of very high surface brightness or low angular momentum.  
Ostriker and Peebles (1973)
have shown that a massive halo will suppress bar instabilities in the disk. 
Thus, it is not surprising that bar-dominated galaxies may exhibit a higher 
fraction of luminous-to-dark matter at their center, to the point of 
possibly being maximal (Quillen \& Frogel 1998, Bottema \& Gerritsen 1998,
Debattista \& Sellwood 1998, Weiner 1998).  

The current analysis together with other recent studies, suggest
a continuum of luminous-to-dark matter content in spiral galaxies
directly proportional to their mean surface brightness (or initial
angular momentum).  LSB and dwarf galaxies appear to be fully DM 
dominated at all radii (Broeils 1992, Zwaan \etal 1995, Bottema 1997, 
C\^ot\'e 1997, Dalcanton \etal 1997, de~Blok \& McGaugh 1997), non-barred 
HSBs are sub-maximal in their inner parts (Bottema 1993, this work),
and bar-dominated galaxies which exhibit, on average, higher surface 
brightnesses than typical Tully-Fisher galaxies may possibly be maximal.  
Following this argument, one may infer that our Milky Way, if it does
harbor a strong bar at its center, 
could indeed be maximal (Sackett 1997).  More explorations of the 
central regions of our Galaxy and the structure of extragalactic 
disks and bulges are desperately needed. 

\section{Conclusion}

We have shown that the TF relation, when cast as
$M_r$~{\it vs.}~$\V22$, is independent of surface brightness
for HSB galaxies. We are lead to interpret this empirical
result as evidence for large quantities of non-luminous matter
in the inner parts  ($R<R_{disk}$) of late-type galaxies.
We have employed simple models for the contraction of the dark matter 
halo by adiabatic infall of baryons (the ``luminous component").
For halos with initially rising rotation curves, this contraction leads to
approximately flat, or featureless, rotation curves, {\sl and} to a TF 
relation that is independent of surface-brightness, if 
$V_{disk}/V_{tot} =0.6 \pm 0.1$ in the mean.
This analysis assumes that the initial halo structure
is uncorrelated with the size of the resulting disk, as suggested by current
cosmological simulations.  Combined with the empirical TF evidence,
this model invalidates the notion of maximal disks for the majority
of HSB late-type spirals.  Our result is independent of the shape
of the luminosity profiles and large variations in our choice of
model parameters.






\bigskip
{\noindent {\bf Acknowledgements}}
The authors would like to thank Roelof Bottema, Alice Quillen and 
Julio Navarro for useful comments, the referee Jerry Sellwood for 
a critical and insightful review, and Adrick Broeils for use of his 
software to produce Figs. 1 \& 2.  



\clearpage


\clearpage

\begin{figure}
\epsscale{1.00}
\plotone{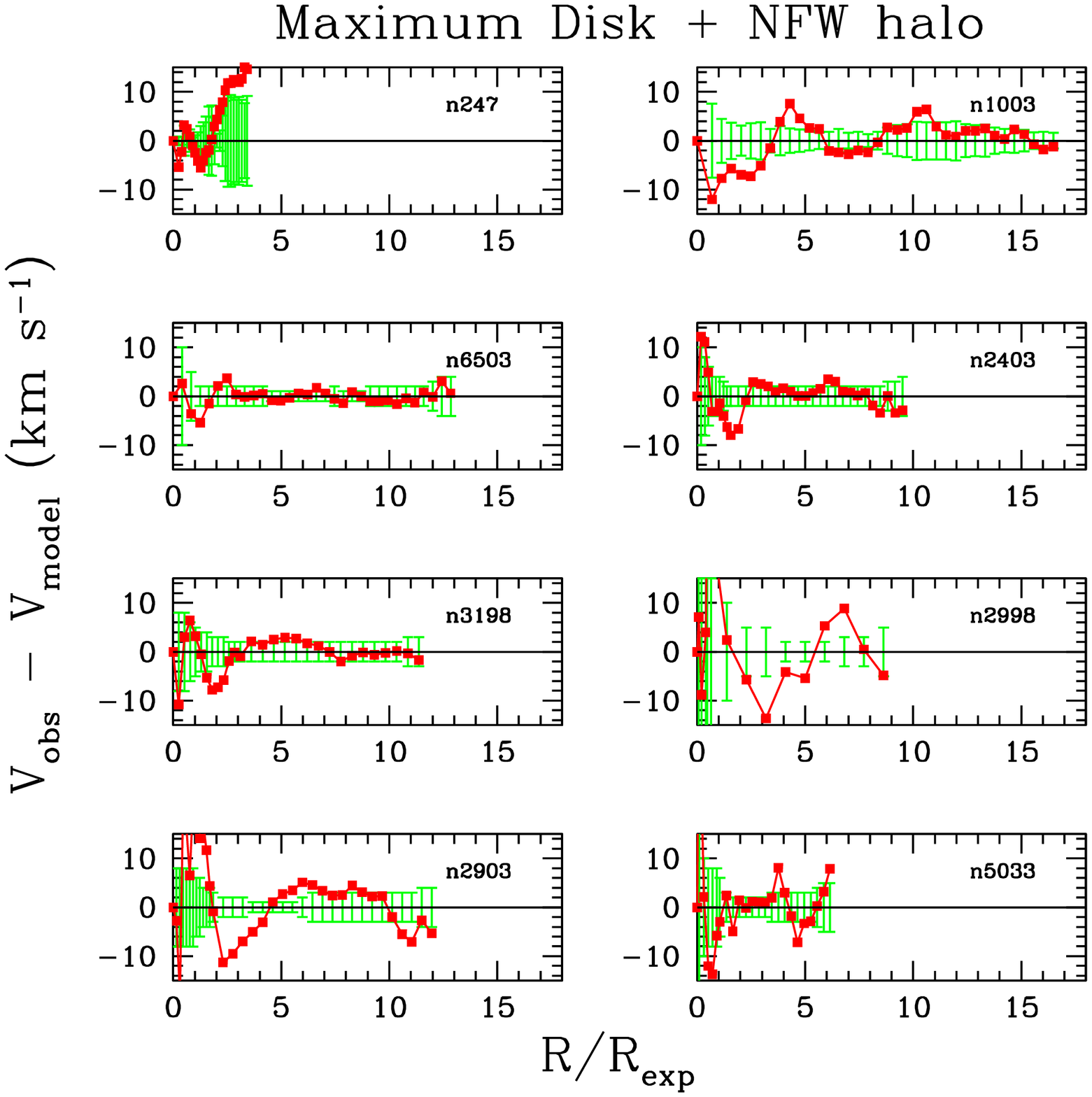}
\caption{Residual differences between maximum disk fits and the observed 
\hi\ rotation curves of Broeils (1992).  Models are shown for all the TF 
galaxies ($V> 200$ \kms) available.  The models contain the maximum amount 
of matter permitted in the disk with $V_{disk}/V_{total} > 0.8$ in all cases. 
This corresponds to a range of $(M/L)_{disk} = 3.5 \pm 1.5$.
}
\end{figure}
\clearpage

\begin{figure}
\epsscale{1.00}
\plotone{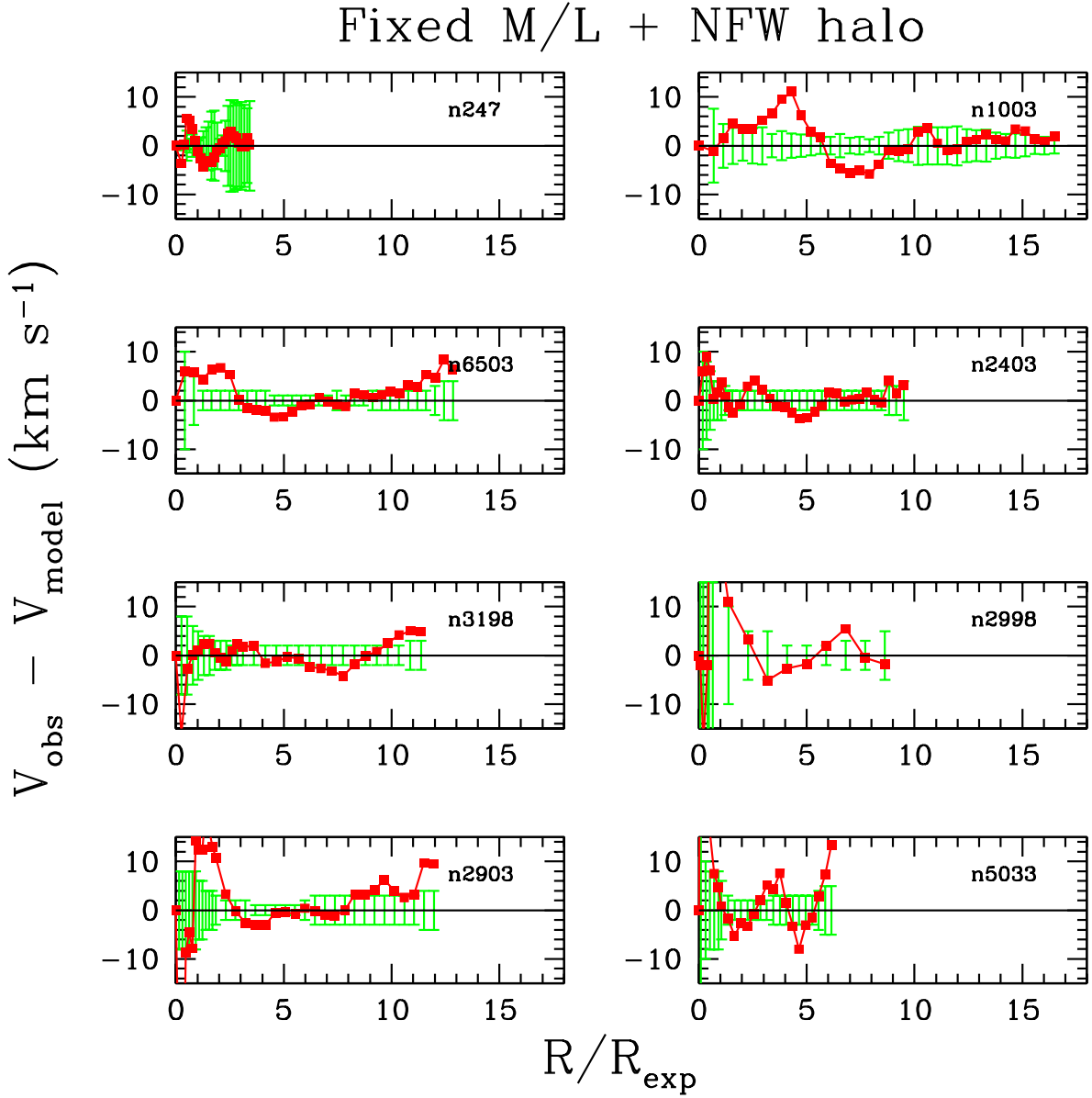}
\caption{Similar data-model residuals as in Fig.~1 but with a lighter disk 
model which obeys $V_{disk}/V_{total} = 0.6$ for each galaxy.  We find a
range $(M/L)_{disk} = 1.5 \pm .7$.  The residuals for the maximum and 
``light'' disks, or \chisqr\ for each rotation curve fit, are 
comparably small and cannot be used to discriminate in favor 
of any one model.  In particular, the specific shape of the NFW 
halo cannot directly constrain the ratio $V_{disk}/V_{total}$ due to
the strong covariance between the concentration parameter of the NFW 
halo and $(M/L)_{disk}$. 
}
\end{figure}
\clearpage

\begin{figure}
\epsscale{0.90}
\vskip -0.3truein
\plotone{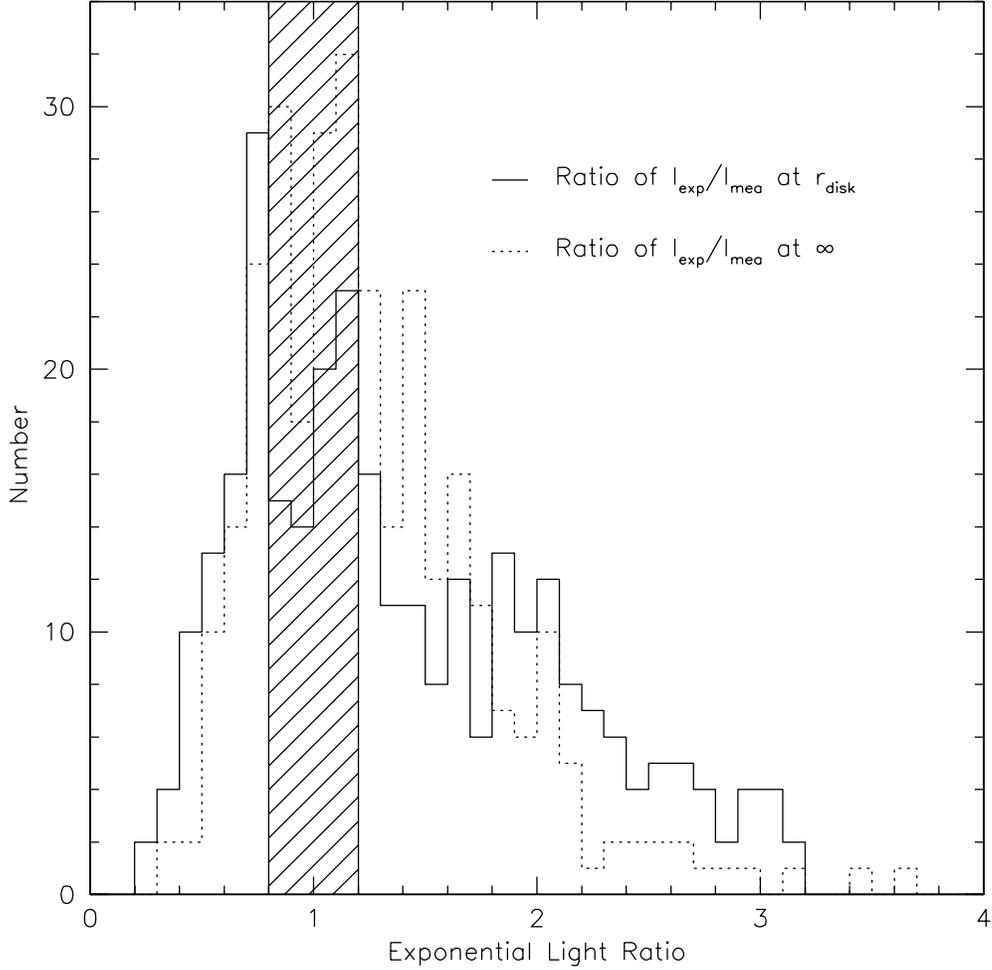}
\caption{Fraction of the light attributed to the fitted exponential 
 disk to the measured light in the galaxy included within $\r22$
 and at ``infinity'' (total extrapolated light).  This histogram is 
 plotted for all CF galaxies.  A significant fraction of late-type 
 disk galaxies deviate from the idealized exponential disk model. 
 We define pure exponential galaxies those which obey 
 $0.8 < \rm{I_{exp}/I_{mea}} < 1.2$ (hashed area).}
\end{figure}
\clearpage

\begin{figure}
\epsscale{0.90}
\plotone{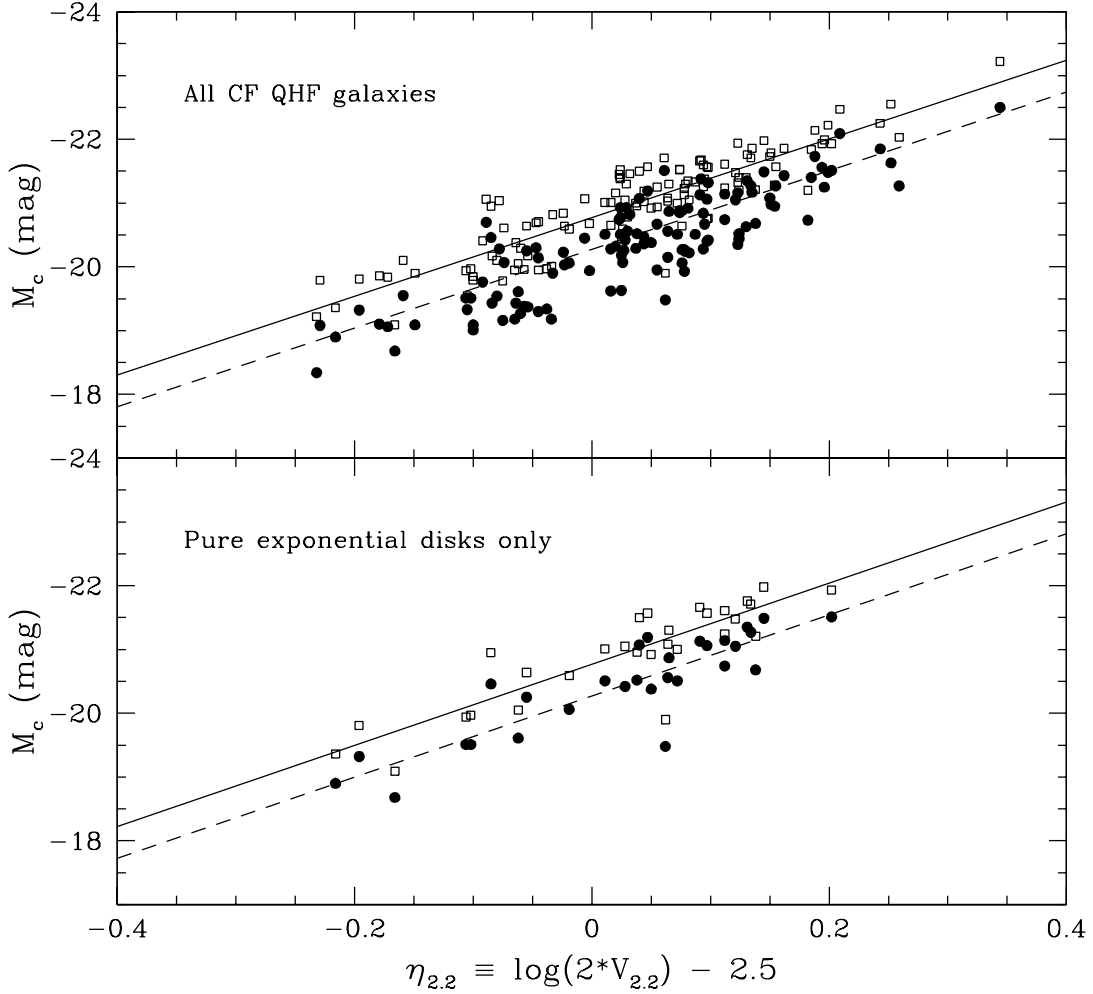}
\caption{Tully-Fisher relations for the CF QHF sample.  Velocities use
 measurements at $v_{2.2}$ for all galaxies.  Squares are plotted for 
 total extrapolated magnitudes, $M_c = M^{tot}_c$, whereas circles 
 use magnitudes within $\r22$, $M_c = M^{2.2}_c$.
 The top figure shows all the CF galaxies and only those deemed ``purely'' 
 exponential are shown at the bottom.  
 The solid line is the Tully-Fisher fit for $M^r_c$ {\it vs} $\eta_{2.2}$
 given in Table 4 of C97 for the Quiet Hubble Flow CF galaxies.  The dashed 
 line is offset from the TF fit by 0\magpoint5.  The magnitude offset between 
 the two data distributions in the top panel is 0\magpoint61.  That 
 offset is identical to 0\magpoint5 in the bottom panel.  The paucity of 
 fast rotators in the bottom panel is explained by the loss of galaxies 
 with big bulges.
}
\end{figure}
\clearpage

\begin{figure}
\epsscale{0.90}
\plotone{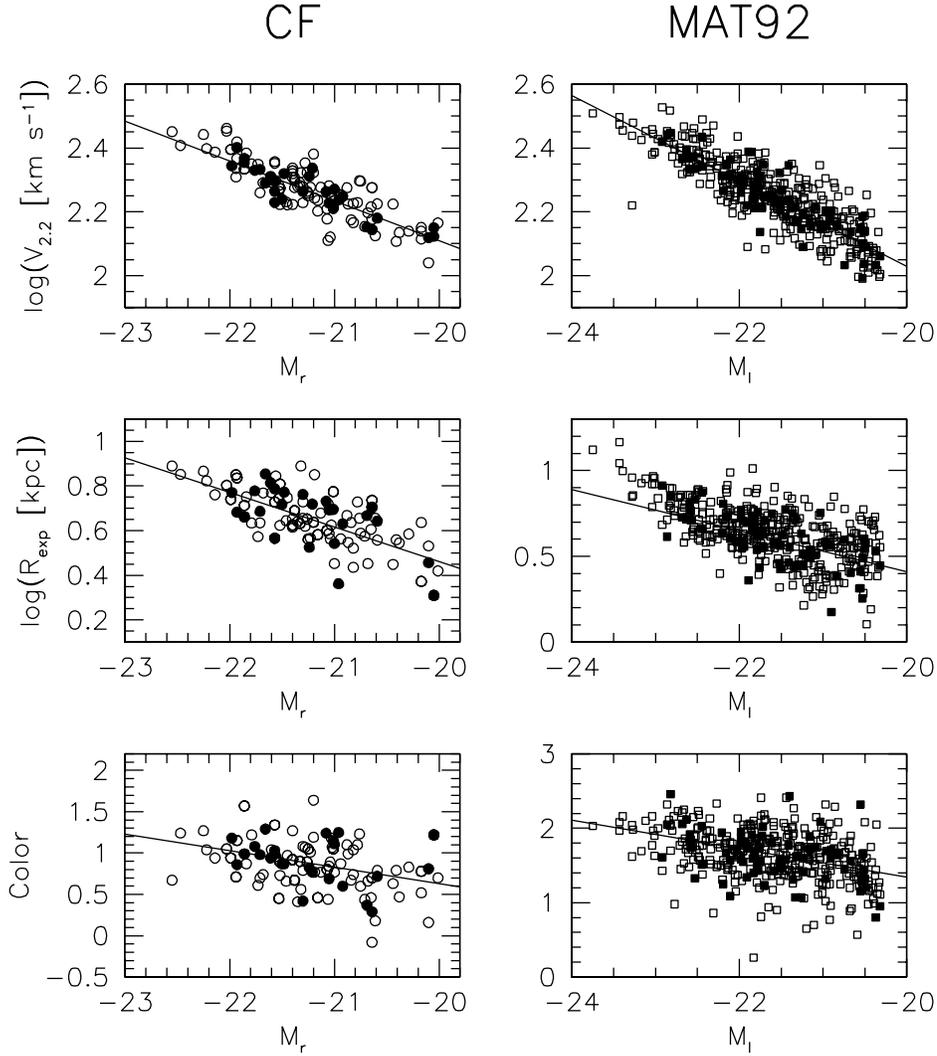}
\caption{Correlations of galaxy's rotational
 velocity, size, and color with absolute magnitude for
 the CF and MAT92 samples.  The color term for each sample 
 corresponds to (B-r) and (B-I) respectively.  
 Open and closed symbols show all the galaxies and
 only the pure exponential disks respectively.
 The solid line in each panel corresponds to a robust fit
 by minimisation of the absolute data$-$model deviations for 
 the pure exponential disks only.  The fits assume equal weights 
 at all points.}
\end{figure}
\clearpage

\begin{figure}
\epsscale{1.00}
\plotone{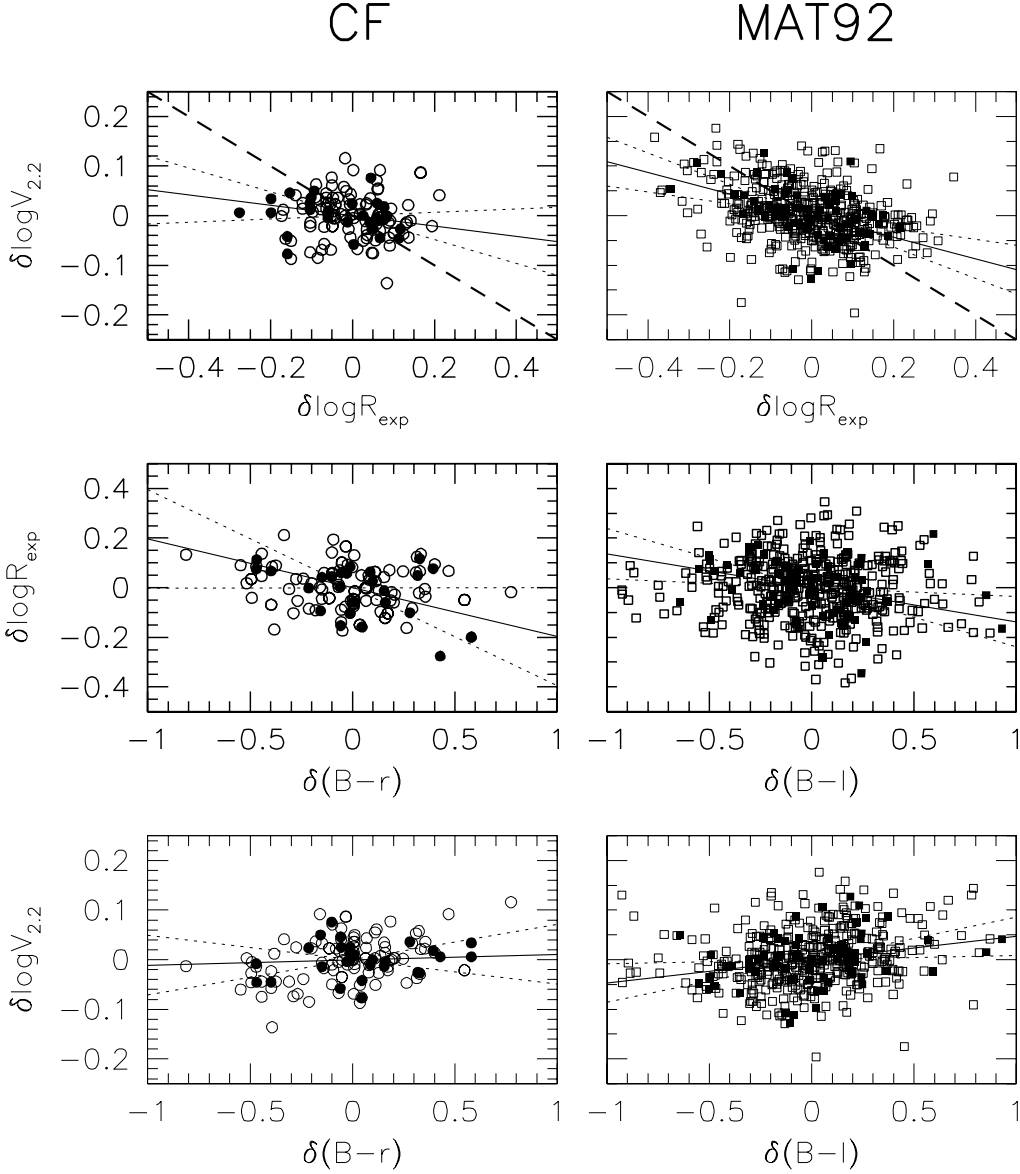}
\caption{Plot of correlation residuals from the mean relation at fixed 
 luminosity for each sample.  The heavy dashed-line in the top two panels 
 show the slope $\partialvr = -0.5$ expected for a maximum disk.  The 
 axis limits are the same for each pair of horizontal panels.  The points
 use the same type convention is Fig.~3.  The solid 
 lines are a best fit to each correlation of residuals for the pure
 exponential disks; the dashed lines represent the statistically acceptable 
 range of these slopes at the 95\% confidence level.  
}
\end{figure}
\clearpage

\begin{figure}
\epsscale{1.00}
\plotone{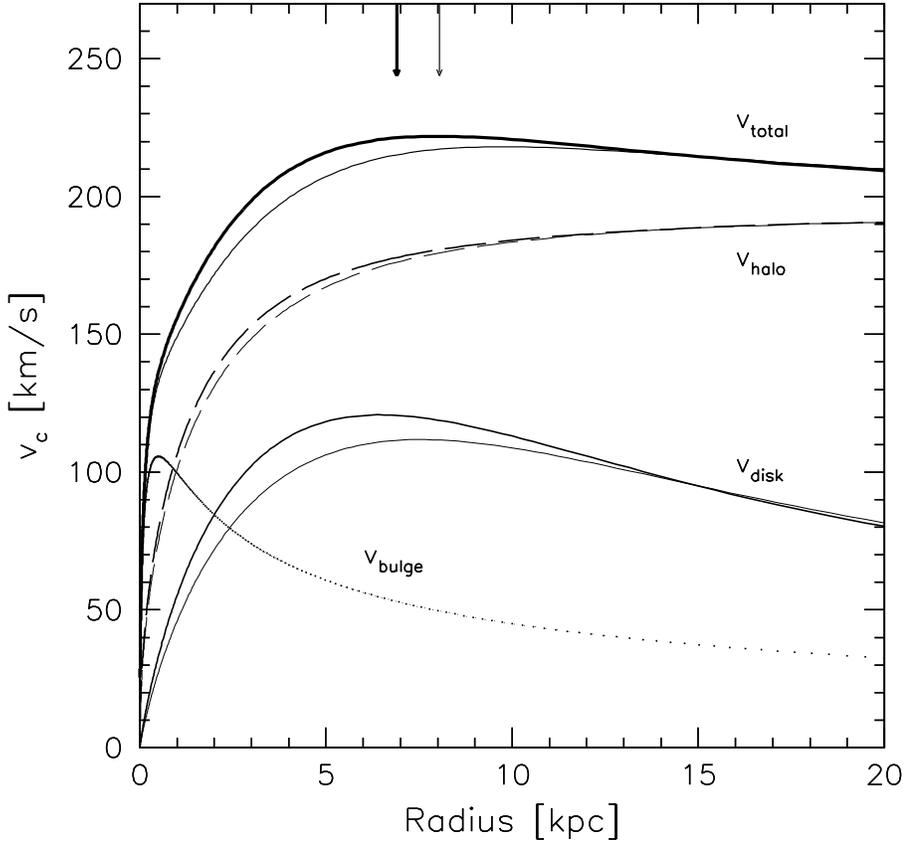}
\caption{
Example of two rotation curves for equal luminosity 
disks of slightly different scale lengths.  These are used to 
calculate $\partialvr$.
The total rotation curve (top solid lines) include contributions
from a bulge, disk and halo.  The halo profile is 
taken from NFW (with $V_{200}=180$\kms) and has been contracted by 
the presence of the disk, assuming adiabatic invariance. 
The thicker set of lines shows the rotation curve for a disk with 
$\Rx = 3$~kpc, while the thinner lines correspond to $\Rx = 3.5$~kpc.
The thick and thin arrows indicate the radii where the velocities 
$\V22$ were determined to compute 
$\partialvr$. 
}
\end{figure}
\clearpage

\begin{figure}
\epsscale{1.0}
\plotone{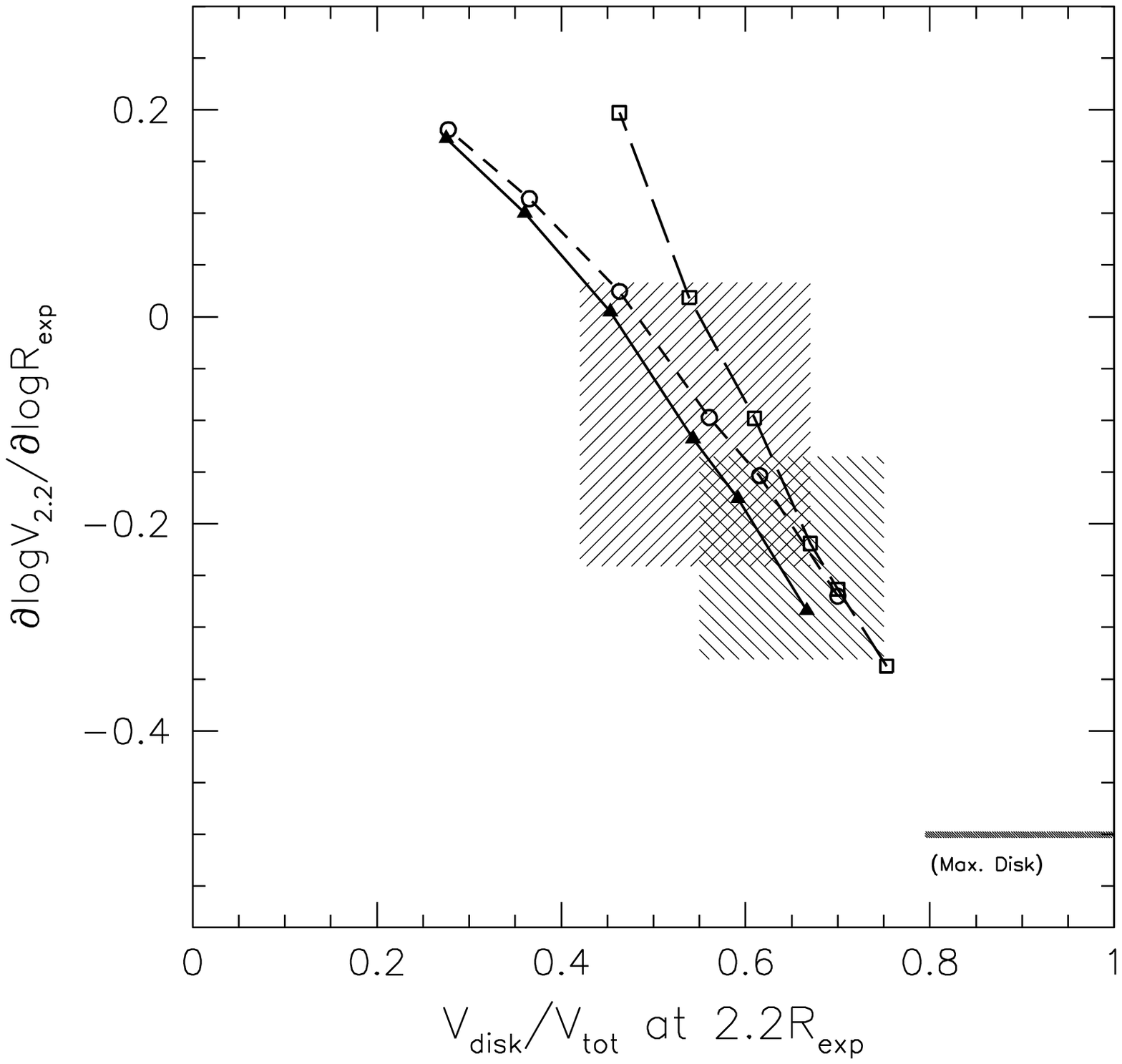}
\caption{
Relation between the luminous and total mass at $2.2\Rx$, 
expressed as $V_{disk}/V_{tot}$ against the gradient $\partialvr$.
The points connected by lines are the result of the numerical
gradient calculations described in \S 3.1. 
For clarity, we show one model (solid line) computed with a NFW halo 
of $V_{200}=200$\kms, a disk of $\Rx=3$~kpc, a bulge of $R_{eff}=500pc$ 
and $M_{bulge}=0.2M_{disk}$.
The disk masses range from $3~10^9M_{\odot}$ to $6~10^{10}M_{\odot}$.
The short-dashed line shows the same case as above but without any bulge.
The long-dashed line shows the case of an isothermal halo ($+$ bulge and disk) 
with an initial core radius of $R_{core}=3\Rx$. The shaded areas indicate
the zone of applicability of the models as constrained by our data (\S 2.3). 
The top shaded box corresponds to limits set by the CF/QHF sample; the lower 
shaded area shows limits obtained with the MAT92 
sample. 
For all disk$-$halo models, the disk must provide on average between 
0.55 and 0.67 of the rotation speed at $2.2\Rx$, in order to be consistent 
with both data samples. 
}
\end{figure}
\clearpage

\begin{figure}
\epsscale{1.0}
\plotone{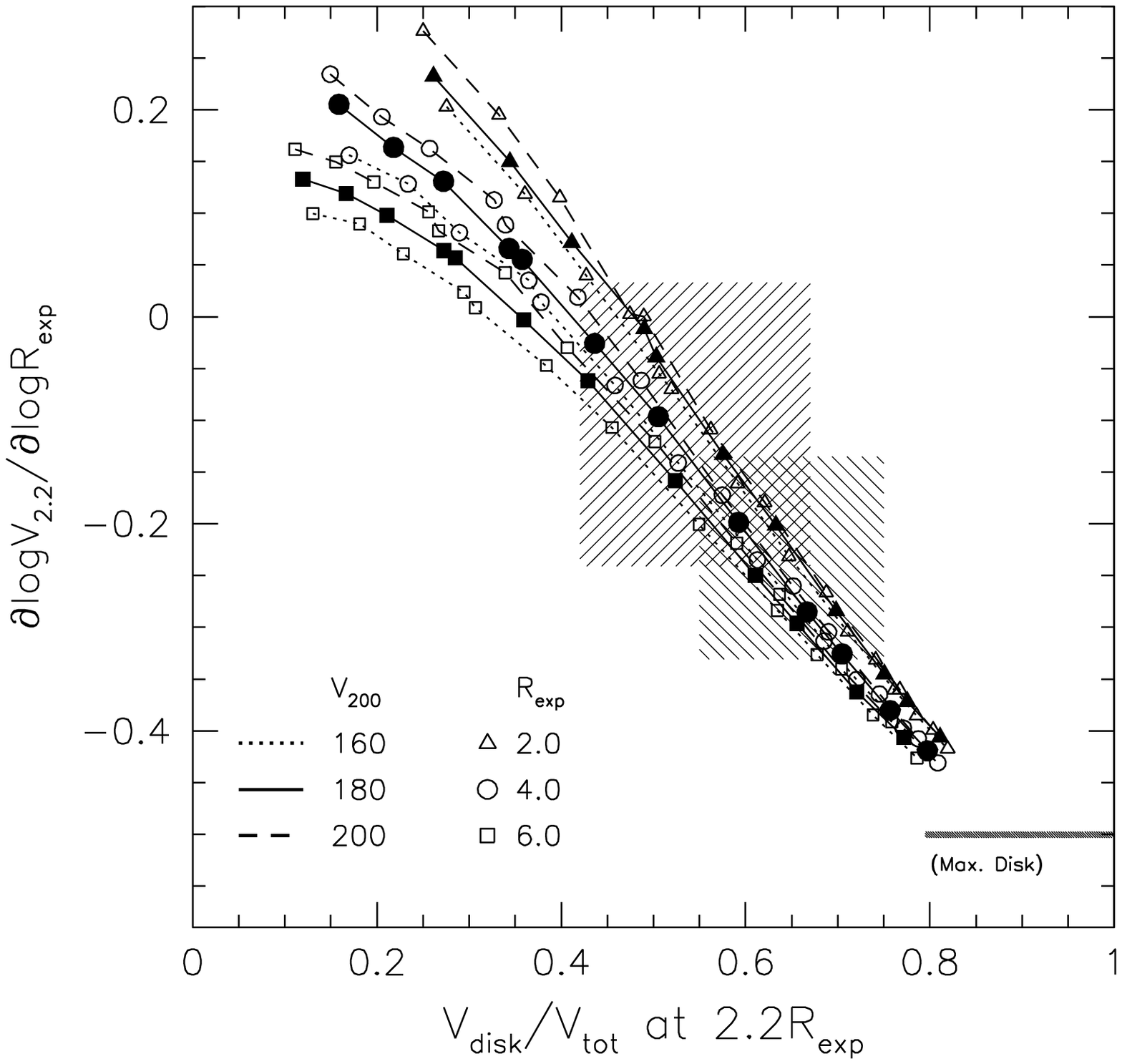}
\caption{
Variation of input parameters $V_{200}$ and $R_{exp}$ on 
model predictions of the gradient $\partialvr$ for any 
$V_{disk}/V_{tot}$.  
We compute galaxy models based on NFW halos with 9 combinations 
of $V_{200}$ and $\Rx$, a bulge of $R_{eff}=500pc$ and $M_{bulge}=0.2M_{disk}$.
The disk masses range from $3~10^9M_{\odot}$ to $6~10^{10}M_{\odot}$.
The filled symbols correspond to the case with $V_{200}=180\kms$.
The hashed areas correspond to confidence limits set by our data.
(see Fig.~8).  This figure shows that galaxies of different stellar
luminosities but similar halo masses do not significantly differ
dynamically.  
}
\end{figure}
\clearpage

\end{document}